\pdfoutput=1 % only if pdf/png/jpg images are used
\documentclass{winnower}

\usepackage{url}
\usepackage{amsmath}
\usepackage[utf8]{inputenc}
\usepackage{caption}
\usepackage{subcaption}
\usepackage{authblk}

\begin{document}

\title{Measurement and simulation of two-phase CO$_2$ cooling for the AFTER electronics of the Micromegas modules for a Large Prototype of a Time Projection Chamber}

\providecommand{\keywords}[1]{\textbf{\textit{Keywords : }} #1}

\author{Deb Sankar Bhattacharya$^{a,b,c}$ \thanks{deb.sankar.bhattacharya@gmail.com}~,
David Atti\'e$^b$, Paul Colas$^b$, Supratik Mukhopadhyay$^a$, Nayana Majumdar$^a$, Sudeb Bhattacharya$^a$, Sandip Sarkar$^a$, Aparajita Bhattacharya$^c$ and Serguei Ganjour$^b$. 

\affil   {$^a$}ANPD, Saha Institute of Nuclear Physics,1/AF Bidhannagar, Kolkata 700064, India{$^a$}.    
\affil   {$^b$}DSM/IRFU/SPP, CEA, Saclay, 91191 Gif Sur Yvette, France {$^b$}.    
\affil   {$^c$}Department of Physics, Jadavpur University, Kolkata 700032, India {$^c$}.   
}

\date{ }
\maketitle

\abstract{  
The readout electronics of a Micromegas (MM) module consume nearly 26 W of electric power, which causes the temperature of electronic board to increase upto $70\,^{\circ}{\rm C}$. Increase in temperature results in damage of electronics. Development of temperature gradient in the Time Projection Chamber (TPC) may affect precise measurement as well. Two-phase CO$_2$ cooling has been applied to remove heat from the MM modules during two test beam experiments at \textit{DESY}, Hamburg. Following the experimental procedure, a comprehensive study of the cooling technique has been accomplished for a single MM module by means of numerical simulation. This paper is focused to discuss the application of two-phase CO$_2$ cooling to keep the temperature below $30\,^{\circ}{\rm C}$ and stabilized within $0.2\,^{\circ}{\rm C}$.}


\section{Introduction}\label{sec:xxx1}
In the International Linear Collider (ILC) \cite{bib1}, electrons and positrons will be collided at center of mass energy upto 1 TeV. The experiments should cover all the physics purposes in the energy range of 240 GeV to 1 TeV. This includes top-quark physics studies, Higgs production and decay, new particle searches, anomalous couplings and Higgs self-couplings. The central tracker in the International Large Detector (ILD) concept \cite{bib2} at the ILC is expected to be a Time Projection Chamber (TPC) \cite{bib3}. Micromegas (MM) detector \cite{bib4} is a suitable candidate for the readout of this TPC.   
Due to the large packing density of the electronic components within a very limited space, the electric power dissipated by the electronics increases the temperature of the detectors. If the heat is not removed within reasonable time, it will cause potential damage to the electronics. Numerical study shows that, due to this heating, gas temperature along the length of the Linear-Collider TPC (LC-TPC) will differ by nearly $35\,^{\circ}{\rm C}$. This may result in non-uniformity of drift velocity of the electrons, hence inaccuracy in measurement. 
Seven Micromegas modules have been tested at the Large Prototype TPC (LP-TPC) \cite{bib5} for ILC at \textit{DESY} with an electron beam of 5 GeV energy. During the beam test, temperature of the electronics has also been monitored. Without cooling, the electronics temperature is found to go as much as upto $70\,^{\circ}{\rm C}$. 

The problem of electronic heating of the Micromegas detectors in large-scale experiments is a serious issue which requires detailed study to estimate probable effects and practical solutions. Use of two-phase CO$_2$ cooling in some industrial fields, as well as in high energy physics experiments like LHCb and AMS-02 \cite{bib6}, shed some light on the problem. In the present work, relevant to the ILD \cite{bib2}, the air cooling system which was used earlier, have been replaced by the two-phase CO$_2$ cooling system. In section 2, the heat generating components of the detector and the effect of heating on a TPC is discussed. 
The advantages of using two-phase CO$_2$ as the coolant are discussed section 3. The experimental details of its application for the cooling of a single MM module and the test results are described in the same section. 
In the next section, a description is given how the same technique is applied for cooling seven MM modules during a beam test. Section 5 is dedicated to the description of a mathematical model of the heating and cooling phenomena inside a detector and solving the problem by numerical calculations. Comparisons of the solutions with the experimental results are also shown here.  

\section{Effect of heating}\label{sec:xxx1.5}
The Micromegas modules are 22 cm $\times$ 17 cm in size. The anode of one such module is segmented in 1728 pads, each having a dimension of 3 mm $\times$ 7 mm. The electronics used for a MM module is called \textit{AFTER} (ASIC For TPC Electronic Readout) \cite{bib7} which runs at 5 volt supply. It consists of six Front End Cards (FEC) and one Front End Mezzanine (FEM).  
The current flowing through each component of the electronics is measured. It shows that the FECs use 19 W power and the FEM takes 3.5 W. Another 3.5 W is consumed by the Field Programmable Gate Array (FPGA). Therefore, the total power dissipation by the entire electronic setup is around 26 W. This power dissipation leads to a rise in the temperature of the electronics upto $65\,^{\circ}{\rm C}$ - $70\,^{\circ}{\rm C}$.

\begin{figure*}[tbp] % figures (and tables) should go top or bottom of
                    % the page where they are first cited or in
                    % subsequent pages
\begin{center}
\includegraphics[width=.6\textwidth]{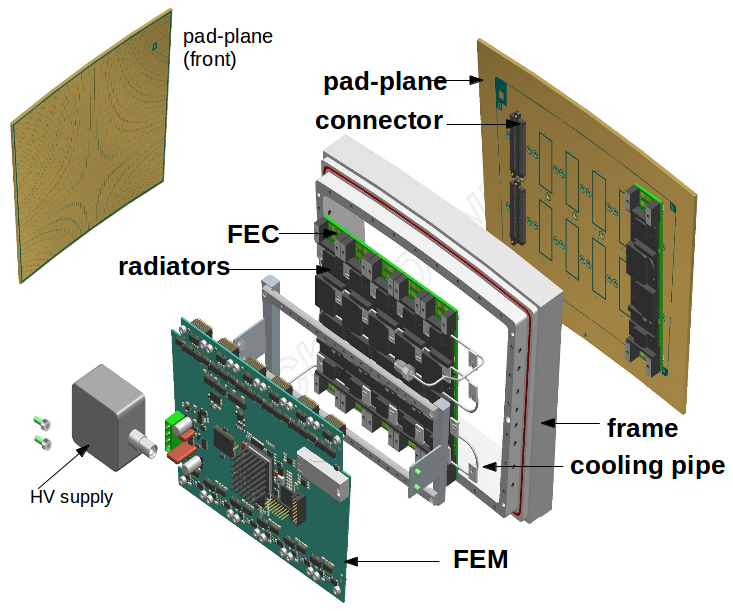}
\caption{Present layout of Micromegas, showing the different components of electronics and the cooling structure}
\label{fig:fig1}
\end{center}
\end{figure*}

\begin{figure}[tbp] % figures (and tables) should go top or bottom of
                    % the page where they are first cited or in
                    % subsequent pages
\centering
\includegraphics[width=.6\textwidth]{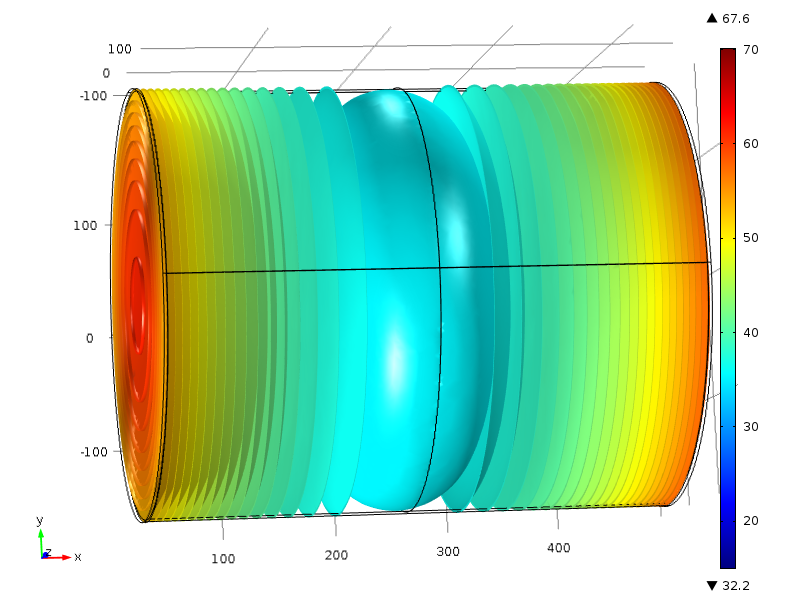}
\caption{The isothermal surfaces show the temperature gradient in Linear Collider TPC after 7 hours of heating. Power consumption at each end plate is taken to be 7.5 kW. The heating is simulated by COMSOL \cite{bib8}.}
\label{fig:fig2}
\end{figure}

\begin{figure}[tbp] % figures (and tables) should go top or bottom of
                    % the page where they are first cited or in
                    % subsequent pages
\centering
\includegraphics[width=.6\textwidth]{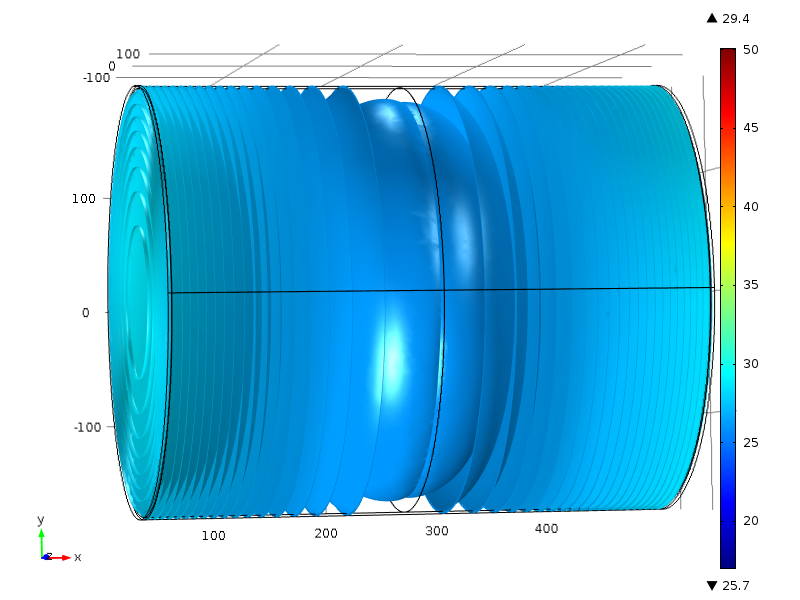}
\caption{The isothermal surfaces show the temperature gradient in Linear Collider TPC after 7 hours of heating. The power consumption is assumed to be reduced by a factor of 10 (750 W for each end plate) with respect to the previous case. This limits the maximum temperature below $30\,^{\circ}{\rm C}$. The heating is simulated by COMSOL \cite{bib8}}
\label{fig:fig3}
\end{figure}

Figure \ref{fig:fig1} shows the layout of the Micromegas detector. The electronic boards (FECs) are connected to the pad-plane (other side of the anode) of the detector by the \emph {connectors}. A mechanical structure holds tightly the FECs on top of the connectors. The metallic pins of the connector can conduct heat to the pad-plane. Excessive heating of the electronics may cause thermal expansion of the mechanical structure on very small scale which in return may give rise to problems like bad connections etc. Certain electronic malfunctioning are also experienced sometimes due to over heating. If the heat is not removed, it will be certainly conducted to the other side of the pad-plane, that is, in the TPC gas. Continuous accumulation of heat on the anodes will cause to warm up the gas at the vicinity and there will be a temperature gradient along the length of the TPC.
A simulation has been carried out on the development of temperature gradient along the Linear Collider TPC (LC-TPC) and it is observed that the difference in temperature can be upto $35\,^{\circ}{\rm C}$. Figure \ref{fig:fig2} and figure \ref{fig:fig3}
show the distribution of isothermal surfaces along the TPC in different conditions.

%Figure ::::::
\begin{figure}[tbp] % figures (and tables) should go top or bottom of
                    % the page where they are first cited or in
                    % subsequent pages
\centering
\includegraphics[width=.6\textwidth]{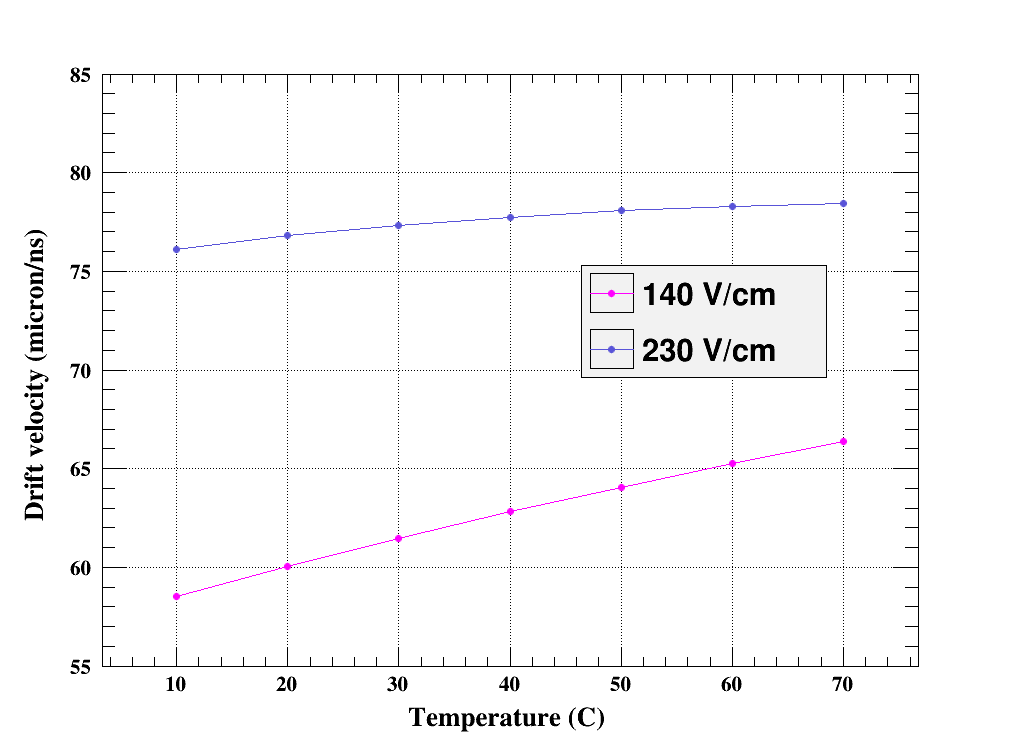}
\caption{Results from simulation with Magboltz \cite{bib9} show the change of drift velocity with temperature for T2K gas. The drift fields are taken as 140 V/cm and 230 V/cm which are standard for T2K gas.}
\label{fig:fig4}
\end{figure}

The simulation is accomplished with COMSOL Multiphysics \cite{bib8} which is based on Finite Element Method. The length of the LC-TPC is taken to be 4.6 m while the diameter is 3.6 m. The common cathode, which is 0.1 mm thick, is at the center. 
The thickness of the field cage and the end-plate of the TPC is taken to be 2 cm. The field-cage and the end-plate are both considered to be made of copper. Considering the present status of the MM electronics for the Large Prototype TPC (LP-TPC), the power consumption is taken to be 7.5 kW for each end-plate. The heat is considered to be uniformly distributed over the entire end-plates. In the simulation, the gas flow is taken to be such that it replaces one TPC volume in 24 hours. As shown in figure \ref{fig:fig2}, the temperature is found to be varying from  $32\,^{\circ}{\rm C}$ to $68\,^{\circ}{\rm C}$ from Cathode to Anode plane of the TPC. Figure \ref{fig:fig3} shows temperature gradient in the TPC when the power consumption is assumed to be reduced by 10 factor with respect to the previous case. The latter considers the possibility of limiting heat generation by suitable means. Now, once temperature gradient is estimated, a simulation is carried out using Magboltz \cite{bib9} within the aforesaid range of temperature for T2K gas under standard drift-fields. The result is shown in figure \ref{fig:fig4}. 

From figure \ref{fig:fig2} and \ref{fig:fig4}, it is fair to conclude that if the heat is not removed, the heat gradient will induce non-uniformity in drift velocity along the length of the TPC. Inhomogeneity in the drift velocity of the electrons is further likely to influence time estimates of the events.
%Figure ::::::::
There could be several options, like limiting heat generation (power-pulsing \cite{bib10}) and removing heat (cooling). A possible way of cooling could be flowing the TPC gas in a temperature controlled way. But it may not be very trivial to maintain the temperature at a constant value and uniformly throughout the TPC while flowing the gas. So, detailed experiment is needed to be carried out in order to adapt this solution. An efficient way of cooling the electronics is to use a coolant flowing through cooling pipes placed on the detector modules. In this report, extensive studies including experiments and simulations on this process are presented.

Few simplifying assumptions have been made in the simulation of LC-TPC: In figure \ref{fig:fig2} and figure \ref{fig:fig3}, the initial and ambient temperature is taken as $25\,^{\circ}{\rm C}$. However, in real case the ambient temperature of the LC-TPC may be even higher by $5\,^{\circ}{\rm C}$-$10\,^{\circ}{\rm C}$. It depends upon the climate and also on the fact that the TPC will be at the center of ILD and will share the heat generated by the other devices as well. The sampling frequency of the electronics for the experiments at LP-TPC is presently set at 25 MHz, that is, time binning of 40 ns. At LC-TPC, where the event generation would be in a much higher rate, the sampling frequency could be even higher to ensure better time reconstruction over a large drift length. Though the exact value is not yet decided, a higher sampling rate will very likely lead to an increase in the power consumption of the electronics. 
\section{Two-phase CO$_2$ cooling for a single MM module}\label{sec:xxx2}
In this section, the usefulness of CO$_2$ cooling and the results are discussed. 

\subsection{Experimental setup}\label{sec:yyy1}
In order to estimate the heating and the effect of cooling, one temperature sensor is plugged in each FEC and the FEM to read temperature at regular time intervals. To apply cooling, the coolant is allowed to flow through a small pipe. The five-fold pipe is equipped with six radiators (figure \ref{fig:fig1}) and the radiators are coupled with each FEC to ensure good thermal contact. The material of the cooling pipe is chosen to be stainless steel, which are essentially non-ferromagnetic to avoid any magnetic induction (so that the setup is ready to be used for the LP-TPC \cite{bib5}). The pipe has standard inner and outer diameters of 0.079 cm and 0.159 cm respectively. Finally, the inlet and outlet of the entire cooling pipe are placed at the two diagonally opposite corners of the module.

\subsection{Why two-phase CO$_2$ cooling}\label{sec:yyy2} 
\begin{figure}[tbp] % figures (and tables) should go top or bottom of
                    % the page where they are first cited or in
                    % subsequent pages
\centering
\includegraphics[width=.7\textwidth]{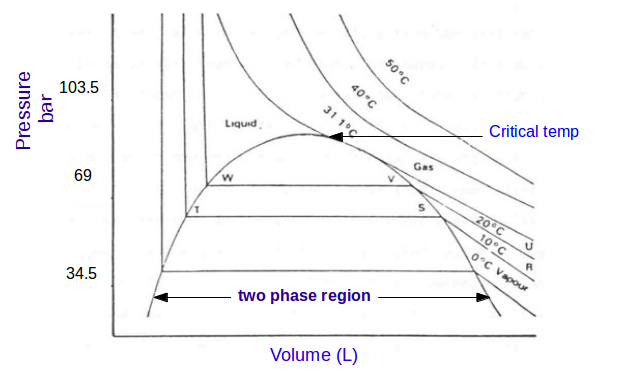}
\caption{Isotherms of CO$_2$ showing the two-phase region and critical point.}
\label{fig:fig5}
\end{figure}

In the beginning, air cooling was tried. However, during the beam test, it was found that the cooling provided by air is not sufficient. This is because, air is not an efficient coolant as it has low thermal conductivity and high viscosity. Two-phase CO$_2$ is considered as an efficient coolant. The principal advantage of using two-phase cooling is that, the heat transfer happens during phase change and so the temperature remains constant. This property allows to maintain a very stable temperature during experiment. The temperature range at which the detector is desired to be cooled is quite below the critical temperature and coinside with the two-phase zone of CO$_2$ (see fig \ref{fig:fig5}).

%Figure :::::::::::::::::::::
\begin{figure}
\begin{subfigure}{.5\textwidth}
  \centering
  \includegraphics[width=1.0\linewidth]{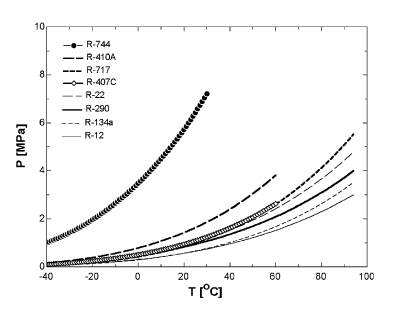}
  \caption{}
  \label{fig:sfig6_1}
\end{subfigure}%
\begin{subfigure}{.5\textwidth}
  \centering
  \includegraphics[width=1.0\linewidth]{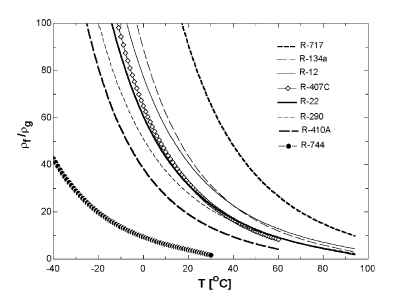}
  \caption{}
  \label{fig:sfig6_2}
\end{subfigure}
\caption{Two useful thermal properties of few commonly used refrigerants. Here R-744 implies CO$_2$. (a) Change of vapor pressure with temperature. (b) Change of the ratio of liquid to vapor density with temperature. The results are taken from \cite{bib11}}
\label{fig:fig6}
\end{figure}

Steeper slope of the vapor pressure curve (see figure \ref{fig:sfig6_1}) helps to change the temperature in smaller steps. It also means a smaller temperature variations against pressure fluctuations. Low ratio of liquid to gas densities results in a more homogeneous two-phase flow (see figure \ref{fig:sfig6_2}). The ratio plays an important role for an evaporator as it determines the flow pattern and hence the heat transfer coefficients \cite{bib11}. The liquid CO$_2$ is well compressible in the range of temperature where the experiment is carried out. Besides that, liquid CO$_2$ has large latent heat which suffices low mass flow rate. Low viscosity provides uniform flow. High compressibility is another interesting property of liquid CO$_2$ which provides high pressure drop which in turn provides the advantage of small tubing.
It should be mentioned in this context, that the cooling pipe is not supposed to be cooled down below $10\,^{\circ}{\rm C}$. Because below $10\,^{\circ}{\rm C}$, ambient water vapor may start to condense around the cooling pipe. If the condensation goes on, the water will flow to the electronics causing damage to the readout of the detector.

\subsection{The cooling system}\label{sec:yyy3} 
\begin{figure}[tbp] % figures (and tables) should go top or bottom of
                    % the page where they are first cited or in
                    % subsequent pages
\centering
\includegraphics[width=.7\linewidth]{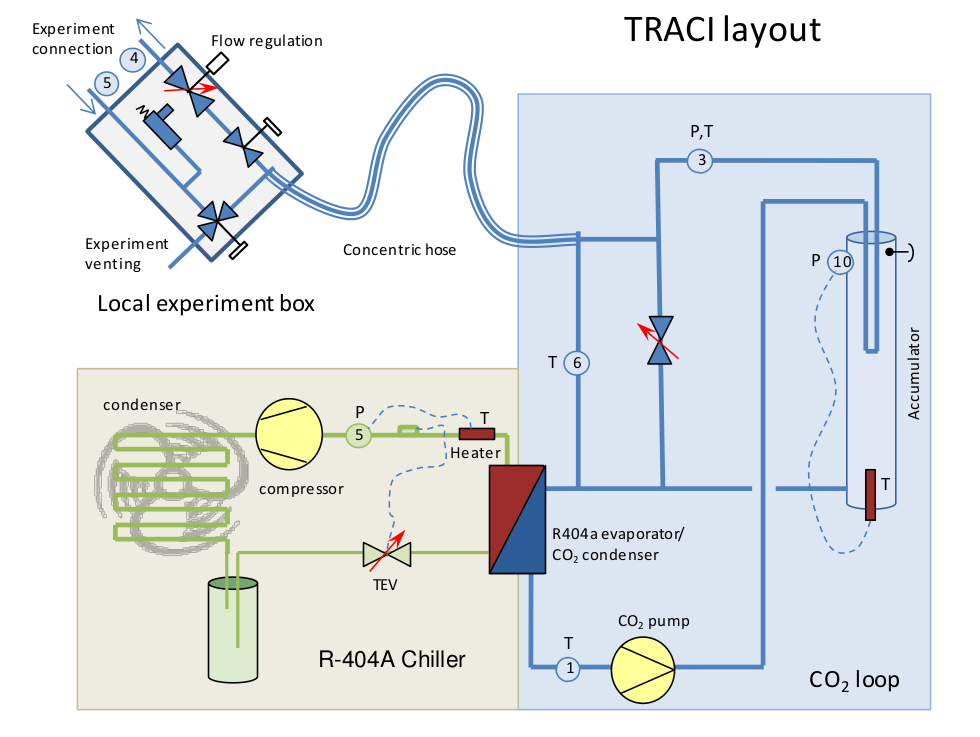}
\caption{The layout of Transportable Refrigeration Apparatus for CO$_2$ Investigation (TRACI)}
\label{fig:fig7}
\end{figure}

The cooling structure, which consists of the cooling pipe and six radiators, is tested with a cooling system \cite{bib12} called TRACI (Transportable Refrigeration Apparatus for CO$_2$ Investigation), at \textit{Nikhef,} Amsterdam in December 2013. The volume of the system is 1.7 l and the CO$_2$ filling is 1 kg. The system consists of chiller, CO$_2$ pump, CO$_2$ condenser, accumulator and a gas distribution box (described in the schematic, figure \ref{fig:fig7}) \cite{bib13}. The distribution (or bypass) box is connected to system by a long concentric hose. The inlets and outlets to the cooling pipes are connected to the bypass box. 
\paragraph* {Features of TRACI:}
\begin{enumerate}
\item It circulates CO$_2$ in close loop and does not throw out CO$_2$ directly to the environment.
\item It can circulate CO$_2$ in a stable pressure difference with a large range of operation.
\item Evaporation point of liquid CO$_2$ can be varied over a long range by adjusting inlet pressure. A calibration chart comes with the system.
\item It provides a long range of flow rate control.
\item Adjusting the inlet and outlet knobs of the bypass box, pressure difference can be changed in desired way.
\item The interlock safety option takes care of pressure difference and automatically shuts down the system if it goes beyond the allowed range.  
\end{enumerate}

Before flowing CO$_2$, the cooling tubes are pumped out for around thirty minutes. In the meantime, the electronic board of the MM module is powered on and it starts getting warm. Finally, CO$_2$ flow is started. The embedded processor board starts taking signals from the temperature sensors and the temperature profile is being written in a file.

\begin{figure}[tbp] % figures (and tables) should go top or bottom of
                    % the page where they are first cited or in
                    % subsequent pages
\centering
\includegraphics[width=.6\textwidth]{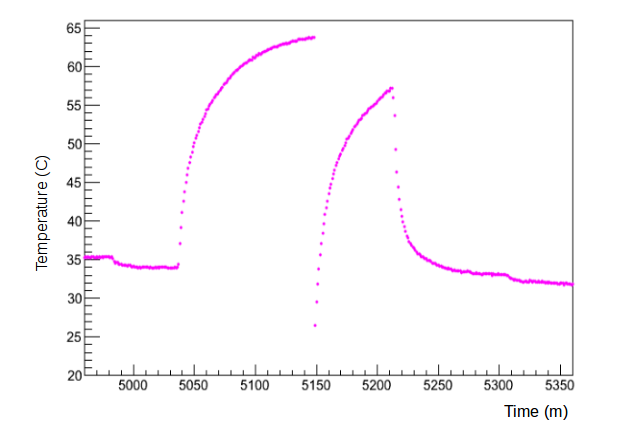}
\caption{Temperature profile of the fifth FEC of the same MM module during test at \textit{Nikhef}}
\label{fig:fig8}
\end{figure}

%\subsection{Test result}\label{sec:yyy4}
\paragraph* {Results:}
Figure \ref{fig:fig8} shows the temperature variation of the fifth FEC. The first flat part of the curve implies that the cooling has been started. Soon after when cooling is stopped, the curve starts to rise sharply and reaches nearly $65\,^{\circ}{\rm C}$. Electronics is then switched off. After few minutes when the electronics is started again, the temperature also starts to rise. Then cooling is circulated and again temperature goes to a stable value. The variation in temperature agrees with the pictures of the pad-plane taken by a thermal camera \cite{bib14,bib15}. This plot contains both temperature rise and fall. Simulation (section 5) has been carried out to study similar temperature variation for one module and the result is compared with this plot.

%%%%% Figure %%%%%%%%
%%%%%%%%%%%%%%%%%

\section{Two-phase CO$_2$ cooling at Large Prototype-TPC}\label{sec:xxx3}  %%%%%%%%%%%%  SECTION-4
\begin{figure}
\begin{subfigure}{.5\textwidth}
  \centering
  \includegraphics[width=.9\linewidth]{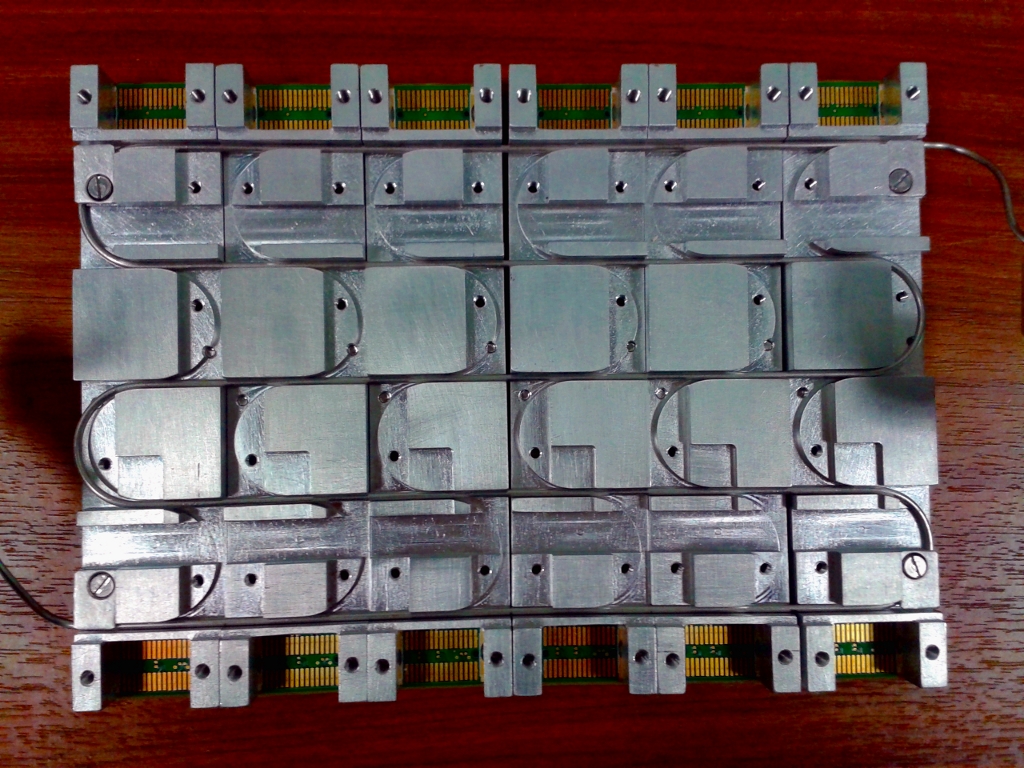}
  \caption{}
  \label{fig:sfig9_1}
\end{subfigure}%
\begin{subfigure}{.5\textwidth}
  \centering
  \includegraphics[width=.8\linewidth]{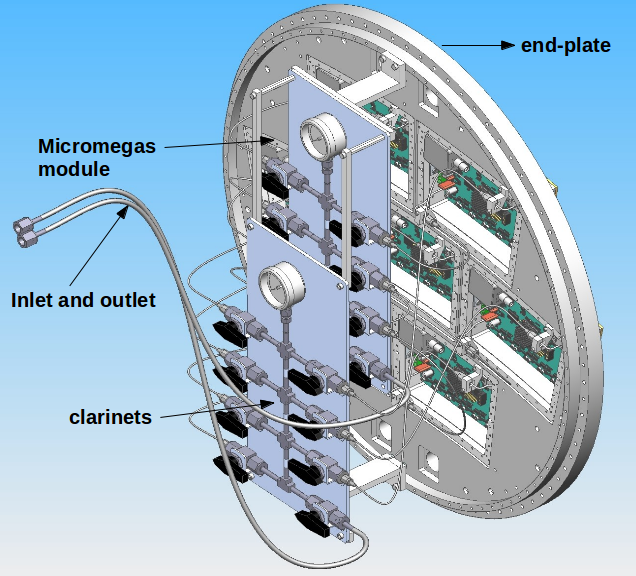}
  \caption{}
  \label{fig:sfig9_2}
\end{subfigure}
\caption{Use of CO$_2$ cooling for all seven modules. (a) radiators with the cooling pipe (b) schematic of the clarinets for CO$_2$ distribution with the LP-TPC end-plate}
\label{fig:fig9}
\end{figure}

An attempt is taken to apply the same cooling technique for all seven MM modules, which are commissioned at the LP-TPC end-plane for test beam \cite{bib15}.
%\subsection{Setup}\label{sec:yyy6}
\paragraph* {Setup:}
After the experiment with a single module, the cooling structure is updated before it is applied for all seven modules at the LP-TPC. The present setup replaced the old air-cooling radiators by aluminum block radiators. The new shape of the radiator with cooling pipe is shown in figure 1 and in figure \ref{fig:sfig9_1}. The pipe is placed in such a way that it may not rest along the length of the FECs but rather in a transverse direction. This makes the cooling uniform.
Seven MM modules are prepared for commissioning at the LP-TPC end-plate. Each of them are equipped with a set of radiators and a cooling pipe. The circulation of CO$_2$ is divided in parallel to all the modules so that the flow could be uniform for all of the modules and also the heat transfer may happen only during phase change which ensures uniform cooling. The distribution of the parallel flow is accomplished with two clarinets. One distributes the inlets while the other is for the outlets. The clarinets are shown in figure \ref{fig:sfig9_2}. Each part of the cooling structure is made up of non-ferromagnetic materials because they would be used 
under 1 T magnetic field. The cooling system is kept far away from the TPC and connected to the bypass box with a long hose. The same setup is applied during test beam in February 2014 and March 2015. It performed consistently each time for more than 80 hours. 

\begin{figure}[tbp] % figures (and tables) should go top or bottom of
                    % the page where they are first cited or in
                    % subsequent pages
\centering
\includegraphics[width=.8\textwidth]{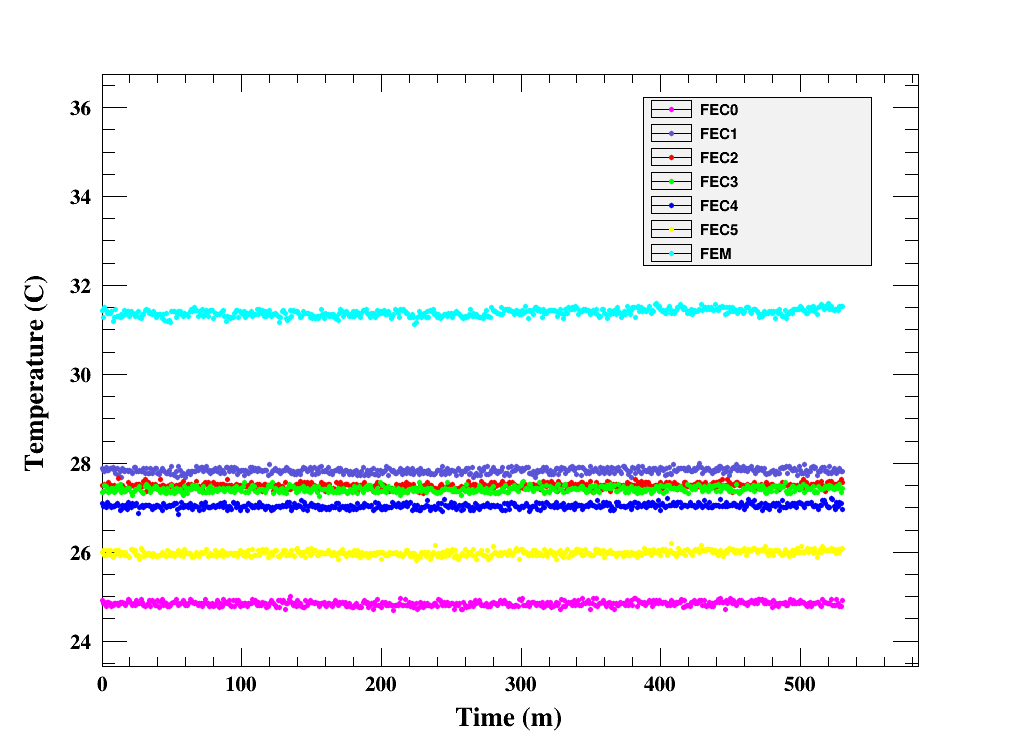}
\caption{Temperature profile of all the FECs and the FEM of a MM module during test beam in March 2015}
\label{fig:fig10}
\end{figure}

%\subsection{Test result}\label{sec:yyy7}
\paragraph* {Test result:}
During the test beam in 2014 and in 2015, the performance of TRACI and the cooling structure is quite satisfactory. The temperature from the temperature sensors from each FECs and the FEMs of all 7 modules is recorded. When plotted, they show a stable temperature over hours during data taking. In figure \ref{fig:fig10}, temperature profile (from test beam March 2015) of all the six FECs and the FEM of a MM module is shown. The figure shows that the setup is successfully implemented to limit the temperature below $28\,^{\circ}{\rm C}$. Temperature of the FEM is seen to be higher than the FECs as it consumes more power. This does not concern too much as the FEM, the outer most component, is relatively far from the pad plane. It may be noted that the temperature is stable within $0.2\,^{\circ}{\rm C}$ during the experiment over days. 

\section{Numerical study for two-phase CO$_2$ cooling}\label{sec:xxx4} 
A generalized study of the cooling process is always desirable because it can help to figure out the influence of the varying parameters and to justify the necessity of the process as well. Moreover, by carrying out a systematic study of the available parameter space, it is also possible to optimize the conceived process in a comprehensive manner. For this reason, numerical simulation of the whole process has always been appreciated upto an achievable limit of precision.

In \textit{SINP}, Kolkata, an attempt to simulate the basic physics of the cooling process has been carried out by making use of COMSOL Multiphysics \cite{bib8}.

\subsection{Mathematical Modeling}\label{sec:yyy8}
The heat transfer processes for a MM module has been mathematically formulated by considering it to be a case of three-dimensional heat flow in a body by the vector method. 
 The MM module has been considered to be a combination of several isotropic bodies having its own thermal conductivity K, density $\rho$ and specific heat $c$. 

The temperature ($\theta$) distribution over such a body having a heat source within it, can be shown to be

\begin{equation}
\label{eq:xxx1}
\rho c {\frac{\partial \theta}{\partial t}}=\vec{\nabla}.(K\vec{\nabla} \theta)+Q
\end{equation}
where Q is the heat produced by the source per second. If there are other sources or sinks of heat they can be accommodated in the equation with appropriate signs.

\subsection{Numerical solution}\label{sec:yyy9}
\begin{figure}[tbp] % figures (and tables) should go top or bottom of
                    % the page where they are first cited or in
                    % subsequent pages
\centering
\includegraphics[width=.6\textwidth]{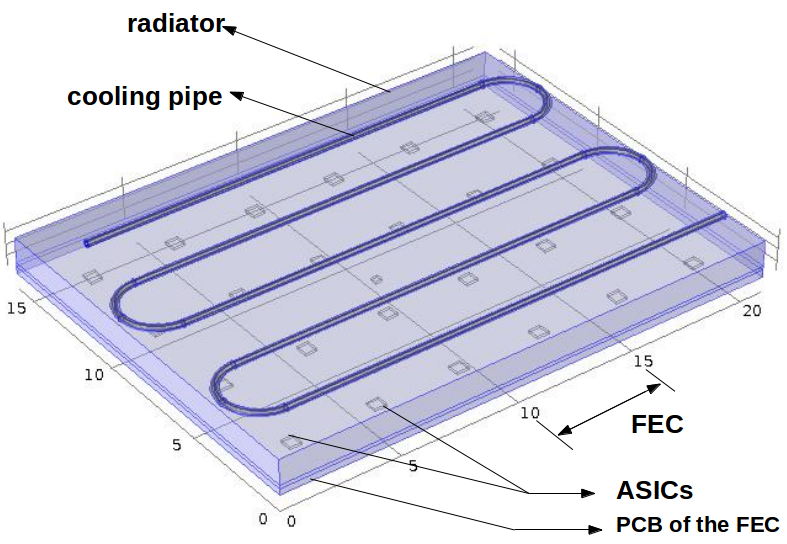}
\caption{The simulated electronics, radiator and cooling pipe (in cm)}
\label{fig:fig11}
\end{figure}

\paragraph* {Model:}
Rise of temperature, effect of cooling etc has been simulated by solving the basic equation described in (\ref{eq:xxx1}) using COMSOL Multiphysics for the entire MM module which consists of isotropic bodies representing the PCB, electronic chips, radiator and the cooling pipes. The three-dimensional virtual geometry, which will be referred as the \emph {model} here-on, has been created such that it closely resembles the real one in all aspects.
The entire simulation is accomplished by using the Heat transfer module within the same software package. The bottom most level of the model geometry is the 21 cm$\times$16 cm PCB. 24 electronic chips of size 5 mm$\times$5 mm each are built in 4 rows and 6 columns on the PCB. Each column represents a FEC. (See figure \ref{fig:fig11})

The spaces between the electronic chips are filled with air. On top of the FECs a thin layer of air (1 mm) is also built. Then a radiator of the same size as of the PCB is built. In the experiment, the radiator is not a continuous piece of metal but a set of 6 identical pieces, one for each FEC. For the sake of simplicity in calculation, the PCB and the radiator both are defined as continuous geometries. The cooling pipes are built on top of the radiator. The pipe is given a five fold shape as done in experiment. It has an outer diameter of 1.58 mm. 
The power consumption by the electronics is already measured to be 26 W. So the entire 26 W power is distributed among the electronic chips. The material property of the different components and the parameters of thermal contact have been defined using COMSOL after carrying out a detailed study on the material and thermal properties of the PCB, the electronic chips and the other parts of the model. The chips has been directly declared as the heat sources. The material and thermal properties of the different components of the model which are needed to solve the concerned equations have been described in the following table. The direction of heat flux is distributed and the convective heat flow is also defined for the air media. Material properties of the fluids applied in the model may vary with temperature and are taken care of COMSOL.

\vspace{2mm}

\begin{tabular}{ |p{3cm}||p{3cm}|p{3cm}|p{3cm}|  }
 \hline
 \multicolumn{4}{|c|}{Physical properties of the different parts of the model} \\
 \hline
 Parts of the Model & Specific-Heat (J/(Kg.K)) & Density (Kg/m$^3$) & Thermal-Conductivity (W/(m.K))\\
 \hline
 PCB   & 385    & 8700   & 400\\
 %\hline
 Electronic chips  & 700 & 2329   & 130\\
 %Air &AL & ALB&  008\\
 Radiator   &475   & 7850  & 44.5\\
 Cooling pipe & 475  & 7000  & 44.5\\
 %Liquid CO$_2$ & AD  & AND   &020\\
 %Gaseous CO$_2$ & AO  & AGO&024\\
 \hline
\end{tabular}

\vspace{1mm}

\paragraph* {Meshing:}
This is an important and critical part of simulation because an optimization between efficiency of computation and accuracy of the solution can be achieved through proper meshing. Fine mesh has been used for geometries with small dimensions and where the solution is expected to vary rapidly (e.g. sharp corners, edges, curves). Necessary mesh-matching has been maintained between elements of significantly different sizes. For instance, the cooling pipe has the diameter of 1.59 mm. Extra fine meshing has been chosen for discretization of the pipe part. Even then, for the inner most layer of the tube, a regular polygonal meshing seems to be too hard to compute and only a quadrilateral meshing is adopted instead. Dealing with very small and relatively large length scales at the same time is a challenging aspect of computation of this kind.
%%%%%%%%  Figure  %%%%%%%%%%%%
%%%%%%%%%%%%%%%%%%%%%%
\begin{figure}
\begin{subfigure}{.5\textwidth}
  \centering
  \includegraphics[width=1.0\linewidth]{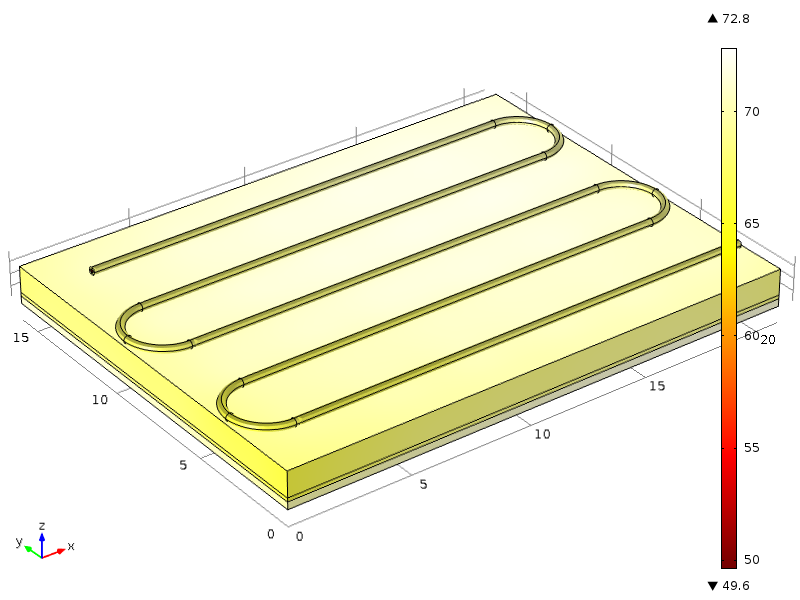}
  \caption{}
  \label{fig:sfig12_1}
\end{subfigure}%
\begin{subfigure}{.5\textwidth}
  \centering
  \includegraphics[width=1.0\linewidth]{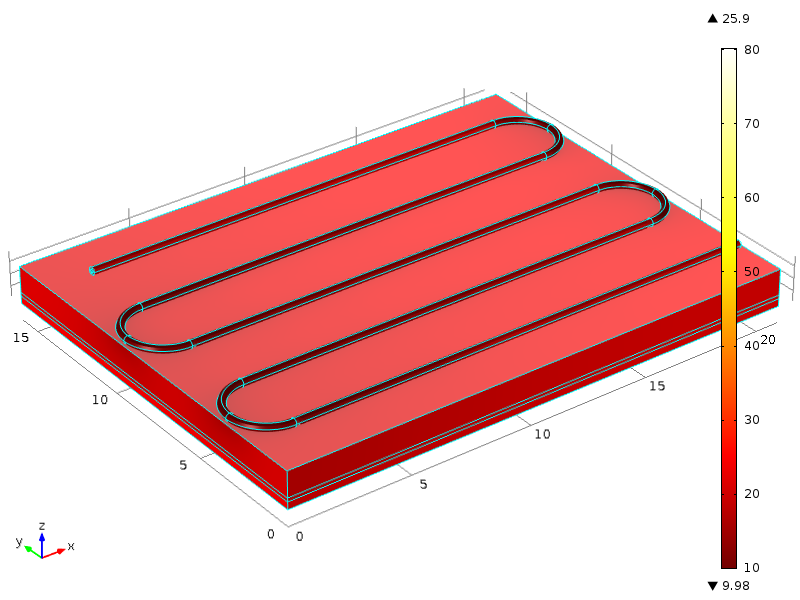}
  \caption{}
  \label{fig:sfig12_2}
\end{subfigure}
\caption{Simulation (dimension is in cm) with COMSOL \cite{bib8} : (a) Temperature of the model after 100 min of heating (b) Temperature of the model 140 min after the cooling is applied. The cooling is done at 10$^o$C and the coolant is flowing at 2 g/sec.}
\label{fig:fig12}
\end{figure}

\paragraph* {Solution:}
Once the meshing is complete the solver in COMSOL solves the following equation to compute heating:
\begin{equation}
\label{eq:xxx3}
\rho c {\frac{\partial \theta}{\partial t}}=\vec{\nabla}.(K\vec{\nabla} T)+Q_{\textit{tot}}
\end{equation} 
Where, \emph T is the temperature, \emph Q$_{\textit{tot}}$ is the total power distributed among the sources.

The initial temperature and power consumption is given as the initial conditions of the problem. At time t = 0 sec the initial temperature is given as room temperature and power consumption is zero. Equation (\ref{eq:xxx3}) calculates the heating up of the model. 

To begin with, the hollow cooling pipe is filled with two-phase CO$_2$. 50\% of the hollow part of the pipe from the center is filled with liquid CO$_2$ where as the rest 50\%, surrounding liquid CO$_2$, is filled with gaseous CO$_2$. Since the temperature is known to remain constant in two-phase cooling, the exact proportion of liquid and gaseous CO$_2$ is not expected to significantly influence the final solution. The common cylindrical surface between liquid and gaseous CO$_2$ is considered to provide the boundary condition for cooling. The surface can be given a constant temperature, for instance 15$^o$C, if cooling is desired at 15$^o$C. Following the experimental situations, CO$_2$ is considered to be flowing inside the tube with a given field velocity. The temperature of the isothermal surface and the field velocity are the parameters which can be tuned in desired way. The temperature at the isothermal surface has been provided as a function of time so that the solver can solve the problem in two parts. Say, for first 100 minutes it solves heating and, when the temperature condition is activated, it starts solving for cooling. For cooling, the COMSOL solves the following equation:
\begin{equation}
\label{eq:xxx4}
\rho c {\frac{\partial \theta}{\partial t}}+\rho c (\vec{u}.\vec{\nabla} T)=\vec{\nabla}.(K\vec{\nabla} T)+Q+Q_{vh}
\end{equation}
where \emph u is the field velocity of CO$_2$ inside the tube, \textit{Q$_{vh}$} is the heat generated due to viscous flow.

\subsection{Result}\label{sec:yyy10}     %%%%%%%    RESULT
\begin{figure}[tbp] % figures (and tables) should go top or bottom of
                    % the page where they are first cited or in
                    % subsequent pages
\centering
\includegraphics[width=.7\textwidth]{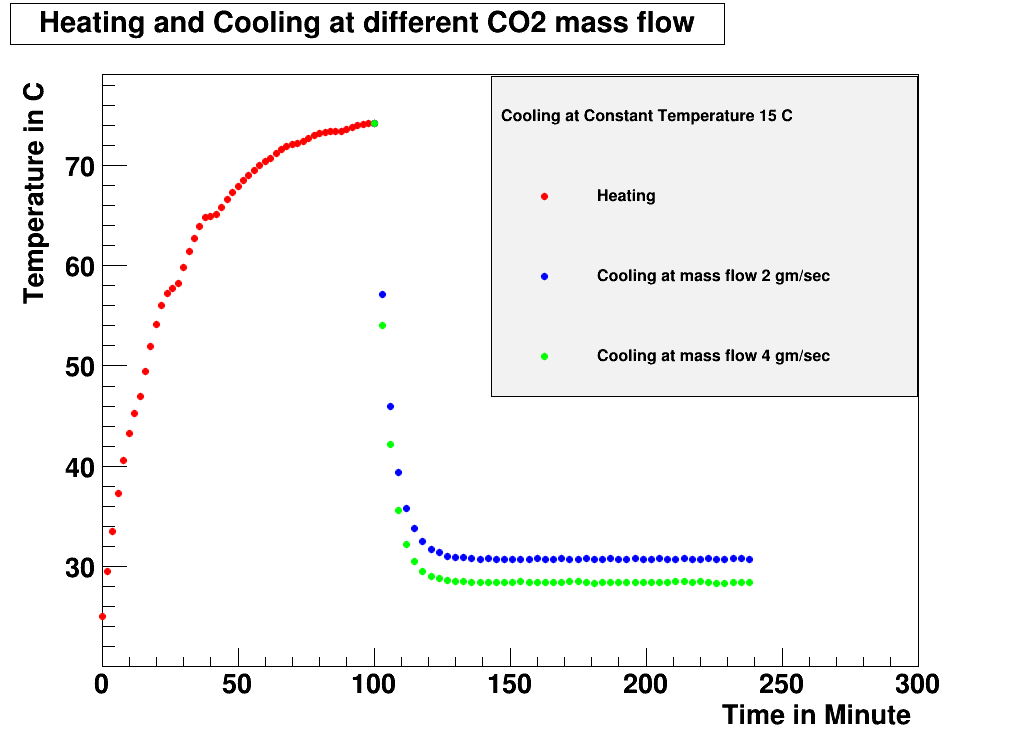}
\caption{Simulation of heating and cooling at two different mass flow rates}
\label{fig:fig13}
\end{figure}

\begin{figure}[tbp] % figures (and tables) should go top or bottom of
                    % the page where they are first cited or in
                    % subsequent pages
\centering
\includegraphics[width=.7\textwidth]{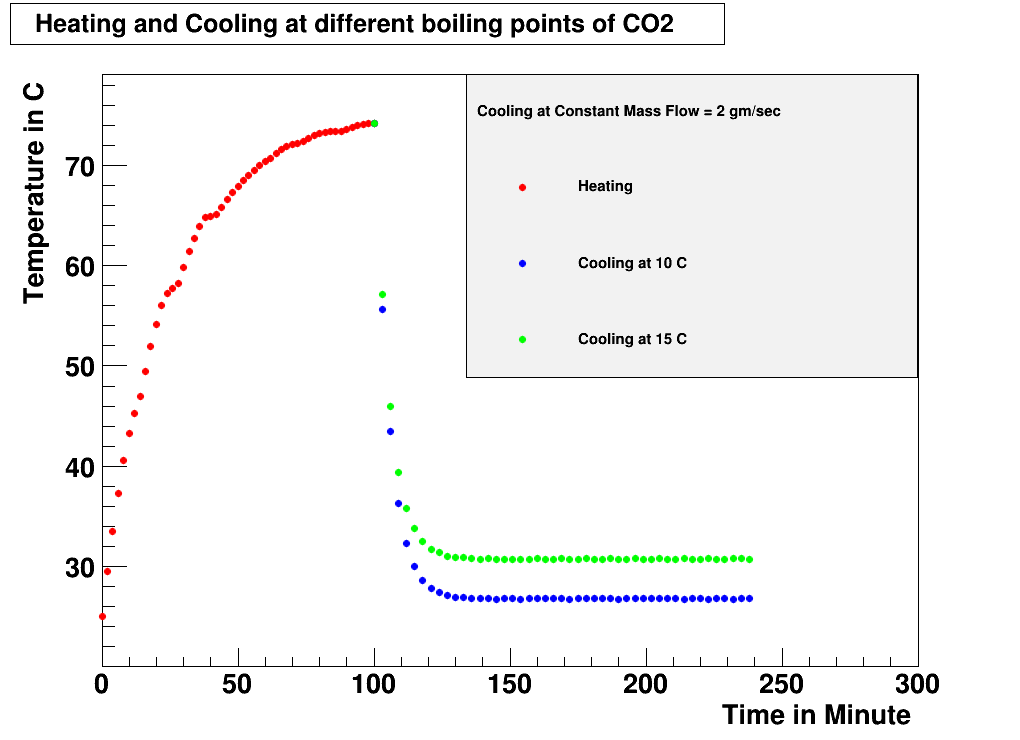}
\caption{Simulation of heating and cooling at two different temperatures}
\label{fig:fig14}
\end{figure}

In simulation, the model is allowed to be heated up for 100 minutes and, next to that, the two-phase CO$_2$ cooling is accomplished for 140 minutes. Figure \ref{fig:sfig12_1} shows the temperature of the simulated model after 100 min of heating and figure \ref{fig:sfig12_2} shows the thermal state of the model 140 min after the cooling is applied. Significant temperature gradients with respect to the environment are observed when no cooling is applied. With the application of cooling, the gradients reduce to much lower values, but are not entirely absent. Detailed investigation on the effects of these gradients is planned in near future. Next, results where the temperature is probed from a point at the center of the model are presented. Figure \ref{fig:fig13} shows the heating and the cooling at two different mass flow rate of liquid CO$_2$. Figure \ref{fig:fig14} describes the heating and the cooling at two different temperature of liquid CO$_2$. In both cases the heating up of the module is done in exactly the same way and hence the curves are identical. In figure \ref{fig:fig13}, the cooling is tested at a constant temperature $15\,^{\circ}{\rm C}$ for two different mass low rates 2 g/sec and 4 g/sec. Increasing the mass flow rate results in lowering the steady-state temperature by $2\,^{\circ}{\rm C}$. This happens because an increase in mass flow rate allows the cooling pipes to absorb even more latent heat for liquid CO$_2$ from the system. Few more extensive modifications on this simulation can be done by considering even precise two-phase fluid flow within the cooling pipes to find a possible saturation limit of lowering steady-state temperature by increasing mass flow rate. In figure \ref{fig:fig14}, the temperature of liquid CO$_2$ is varied at $15\,^{\circ}{\rm C}$ and $10\,^{\circ}{\rm C}$ for a constant mass flow rate of 2 g/sec. Lowering the temperature of liquid CO$_2$ helps to lower the steady-state temperature by around $5\,^{\circ}{\rm C}$.

The experimental heating curve (figure \ref{fig:fig8}) tends to rise $65\,^{\circ}{\rm C}$ in 100 min (before the electronics was switched off) whereas the simulated heating curve is found to rise around $75\,^{\circ}{\rm C}$ in the same span of time. The small difference in simulation may be because of the fact that the whole MM module is not simulated and the geometry of the model is a bit simplified. The only reason of this simplification is to reduce computation load. The cooling curves from both experiment and simulation are very close to $30\,^{\circ}{\rm C}$.   

\section{Conclusion and Plans}\label{sec:xxx5}
In pursuance of removing heat from the back end of the MM detectors, the idea of two-phase CO$_2$ cooling has been first applied on a single MM module. The experiment brought an assurance to solve the problem in an efficient way. The same procedure has been implemented on seven MM modules while they have been commissioned at LP-TPC \cite{bib5} for test beam experiment. The result of cooling is very satisfactory as the temperature have been reduced below $28\,^{\circ}{\rm C}$ and it is stable within $0.2\,^{\circ}{\rm C}$ over days. This efficient cooling has allowed to run the test beam very smoothly. However, there are scopes to improve the setup to make better and uniform thermal contact so as to remove temperature differences between the FECs. Though the FEM is relatively distant from the pad-plane and have less contribution to heat up TPC gas, more investigation can be done to reduce FEM temperature.

The simulation of the cooling process inside a MM module matches the experimental result closely and hence can be useful to study different aspect of this process. After the experimental and numerical studies it may be considered to have a safe solution for cooling of Micromegas electronics for ILD \cite{bib2}. Subjects like better thermal contact, more optimized configuration of the radiator and the cooling pipes can be estimated with the help of the simulated model as well. There is a chance that the effect of power pulsing on heat generation can be estimated from the simulation. The simulated model, with suitable modifications, may be used to optimize cooling in some other kind of detector electronics.

\acknowledgments

We express our gratitude to \textit{KEK}, Japan for financing the cooling system, TRACI. We sincerely acknowledge Jan Timmermans and Bart Verlaat of \textit{Nikhef}, Amsterdam, for their overall assistance and communications. We are also thankful to Marc Riallot of \textit{CEA}, Saclay. We would like to acknowledge Ralf Diener at \textit{DESY}, Hamburg. Our sincere thanks to Purba Bhattacharya, Abhik Jash and Rajani Raman of \textit{SINP}, Kolkata. We express our gratitude to CEFIPRA/IFCPAR (Project 4304-01) and AIDA for their partial fundings.

\bibliographystyle{ieeetr}

% \bibliography{Co2}{}

\end{document}